\documentclass[12pt]{article}

\usepackage{amsmath}
\usepackage{amssymb}

\newcommand{\set}[1]{{\mathbb{#1}}}
\newcommand{\one}{\mbox{\tt 1}\hspace{-0.057 in}\mbox{\tt l}}

\begin{document}

\title{Hypothesis elimination on a quantum 
computer\footnote{in Proceedings of the
                  7th International Conference on
                                  Quantum Communication,
                  Measurement and Computing (QCMC'04),
                  edited by Stephen M. Barnett 
                                  (AIP Press, Melville, NY, 2004).\vspace{5mm} }}

\author{Andrei N. Soklakov\footnote{E-mail: a.soklakov@rhul.ac.uk} \ and 
R\"udiger Schack\footnote{E-mail: r.schack@rhul.ac.uk} \\
{\it Department of Mathematics, Royal Holloway, University of London} \\
{\it Egham, Surrey, TW20 0EX, UK}
}

\date{3 December 2004}

\maketitle

\begin{abstract}
Hypothesis elimination is a special case of Bayesian updating, where each 
piece of new data rules out a set of prior hypotheses. We describe how to use
Grover's algorithm to perform hypothesis elimination for a class of probability
distributions encoded on a register of qubits, and establish a lower bound
on the required computational resources.
\end{abstract}

\section{Introduction}

In the standard setting of Bayesian inference one starts from a set of
hypotheses $\set{H}=\{h\}$ and a set of possible data $\set{D}=\{d\}$.
Hypotheses and data are connected via conditional probabilities
$p(d|h)$, known as the {\em model}. Given a prior
probability distribution $p(h)$, the updated probability $p(h|d)$ of the
hypothesis $h$ to be true given that the piece of data $d$ was observed is
obtained from Bayes's rule, \cite{Bernardo1994}
\begin{equation}        \label{Bayes}
p(h|d)=\frac{p(d|h)p(h)}{\sum_{h\in\set{H}}p(d|h)p(h)} \;.
\end{equation}
In this paper, we consider the problem of hypothesis elimination, which is a special
case of Bayesian updating where the model is of the form
\begin{equation}    \label{eq:model}
p(d|h)=\left\{\begin{array}{ll}
        0 & \mbox{if} \; h \; \mbox{is ruled out by} \; d \,,\cr
                c_h & \mbox{otherwise.}
       \end{array}\right.
\end{equation}
The positive constant $c_h$ does not depend on $d$ and is determined by
normalization. We assume that there is a finite number, $N$, of hypotheses,
which we label 0 to $N-1$, i.e., $\set{H}=\{0,1,\dots,N-1\}$. Furthermore, we
assume that the prior $p(h)$ has been obtained by hypothesis elimination from
an initial uniform prior distribution on $\set H$.

There are several possible ways of encoding a probability distribution $p(h)$ on a
quantum register. Here we represent $p(h)$ by the state
\begin{equation}
|\psi\rangle=\sum_{h=0}^{N-1} \sqrt{p(h)}\; |h\rangle  \;,
\end{equation}
where $|h\rangle$ are  computational basis states \cite{NielsenChuang} 
of a register formed of $\lceil\log_2N\rceil$ qubits.

To be specific, we assume that the prior $p(h)$ has been obtained by $k-1$
hypothesis elimination steps ($k>1$). We thus assume that the prior is given
in the form of a sequence of {\it oracles\/} $o_1,\ldots,o_{k-1}$, where
\begin{equation}
o_j(h)=\left\{\begin{array}{ll}
        0 & \mbox{if} \; h \; \mbox{is ruled by the data in step} \; j \,,\cr
        1 & \mbox{otherwise.}
       \end{array}\right.
\end{equation}
Likewise, we assume that the model Eq.~(\ref{eq:model}) is given as an oracle $o_k$,
where $o_k(h)=c^{-1}_hp(d|h)$. 
The sequence of oracles $o_j$ gives rise to a new sequence
$O_1,\ldots,O_k$ defined by
\begin{equation}
O_j(h)=\left\{\begin{array}{ll}
              1& {\rm if\ } \forall i\leq j:\ o_i(h)=1\,,\cr
                          0& {\rm otherwise}\,.
              \end{array}
       \right.
\end{equation}
For each $O_j$, we define the set of solutions, $\Omega_j=\{h:O_j(h)=1\}$, and
denote by $M_j$  the corresponding number of solutions,
\begin{equation}
M_j=\sum_{h=0}^{N-1}O_j(h)\;.
\end{equation}
For $j\in\{k-1,k\}$, we now define the states
\begin{equation}
|\Psi_j\rangle=\frac{1}{\sqrt{M_j}}\sum_{h\in\Omega_j}|h\rangle\,.
\end{equation}
The state $|\Psi_{k-1}\rangle$ encodes the prior $p(h)$, and $|\Psi_k\rangle$
encodes the posterior $p(h|d)$. 

\section{Quantum hypothesis elimination}

The problem of hypothesis elimination now
takes the following form: Given a quantum register in the prior state
$|\Psi_{k-1}\rangle$ and given the oracles $O_1,\ldots,O_k$, transform the 
register state into the posterior state $|\Psi_{k}\rangle$. One can use Grover's
algorithm \cite{Grover1997} to solve this problem as follows. 

Define a quantum oracle corresponding to $O_k$ via
\begin{equation}
\hat{O}_k|h\rangle=(-1)^{O_k(h)}|h\rangle\,.
\end{equation} 
There are standard techniques~\cite{NielsenChuang} to implement $\hat{O}_k$ in
the form of a quantum circuit.
The Grover operator $\hat{G}(O_k)$ associated with the oracle is then defined as
\begin{equation} \label{GroverOperator}
\hat{G}(O_k)=\big(2|\Psi_0\rangle\langle\Psi_0|-\one\big)\hat{O}_k\,,
\end{equation}
where $\one$ is the identity operator and
\begin{equation}
|\Psi_0\rangle=\frac{1}{\sqrt{N}}\sum_{x=0}^{N-1}|x\rangle
\end{equation}
is the
equal superposition state.
The posterior state $|\Psi_k\rangle$ can now be prepared by repeated
application of the Grover operator $\hat{G}(O_k)$ to the equal superposition
state $|\Psi_0\rangle$. This requires $(\pi/4)\sqrt{N/M_k}$ calls of the
oracle $\hat{O}_k$.

Notice that the hypothesis elimination algorithm outlined above makes no direct use 
of the prior state $|\Psi_{k-1}\rangle$. This raises the following question: 
Is it possible to reduce the number of Grover iterations (and therefore oracle
calls) required to prepare $|\Psi_k\rangle$ by starting from the prior state
$|\Psi_{k-1}\rangle$ instead of the equal superposition state
$|\Psi_0\rangle$? In other words, can one make use of the computational effort
that went into preparing the prior state $|\Psi_{k-1}\rangle$ in order to 
obtain the posterior state $|\Psi_k\rangle$ more efficiently?
As already suggested by the results in Ref.~\cite{Biham1999}, 
the answer to this question is negative. Here we prove the 
following result.

Consider the family of oracles $\set{O}$ that consists of $O_k$, $O_{k-1}$ and all
possible combinations of $O_k$ and $O_{k-1}$ 
(see Eq~(\ref{family}) for the precise definition of this family).  
Now consider all possible
algorithms that consist of applying the corresponding Grover operators
\begin{equation}
{\mathcal{A}}_{abc\dots}=\dots\hat{G}(O_c)\hat{G}(O_b)\hat{G}(O_a)\,,\ \ \ \ \ 
O_a,O_b,O_c,\dots\in\set{O}\,.
\end{equation}
Then, in the limit of large $N$, any algorithm ${\mathcal{A}}_{abc\dots}$
requires at least $(\sqrt{2}/8)\sqrt{{N}/{M_k}}$
oracle calls from the above family to convert $|\Psi_{k-1}\rangle$ into
$|\Psi_k\rangle$. In other words, making direct use of the prior state
$|\Psi_{k-1}\rangle$ does not improve the asymptotic cost of $O(\sqrt{N/M_k})$
oracles calls required to prepare $|\Psi_k\rangle$.

\section{Proof}

Consider an oracle $O$ which accepts $M$ out of the
total $N$ hypotheses $h$:
\begin{equation}  \label{firstEquation}
\sum_{h=0}^{N-1}O(h)=M\,.
\end{equation}
We shall call such
hypotheses {\em good}, as opposed to {\em bad\/} 
hypotheses that are rejected by the oracle. 
Using different notation \cite{soksch1} for the 
amplitudes of good and bad hypotheses,
we have that after $t$ consecutive
applications of the Grover operator 
$\hat{G}(O)$ an arbitrary quantum state
\begin{equation} \label{initial}
|\Psi^{\rm ini}\rangle= \sum_{{\rm good\ }h} g^{\rm ini}_h|h\rangle
                           +\sum_{{\rm bad\ }h} b^{\rm ini}_h|h\rangle
\end{equation}
is transformed into
\begin{equation}
|\Psi^{\rm fin}\rangle=\hat{G}^{\, t}(O)|\Psi^{\rm ini}\rangle
=\sum_{{\rm good\ }h} g^{\rm fin}_h|h\rangle
                           +\sum_{{\rm bad\ }h} b^{\rm fin}_h|h\rangle.
\end{equation}
Let $\bar{g}^{\rm ini}$ and $\bar{b}^{\rm ini}$ be the 
averages of the initial
amplitudes corresponding to the good and the bad 
hypotheses respectively:
\begin{equation}
\bar{g}^{\rm ini}
=\frac{1}{M}\sum_{{\rm good\ }h}g^{\rm ini}_h\,,\hspace*{2cm}
\bar{b}^{\rm ini}=\frac{1}{N-M}\sum_{{\rm bad\ }h}b^{\rm ini}_h\,,
\end{equation}
and similarly for the final amplitudes
\begin{equation}
\bar{g}^{\rm fin}
=\frac{1}{M}\sum_{{\rm good\ }h}g^{\rm fin}_h\,,\hspace*{2cm}
\bar{b}^{\rm fin}=\frac{1}{N-M}\sum_{{\rm bad\ }h}b^{\rm fin}_h\,,
\end{equation}
Let us also define
\begin{equation}
\Delta g^{\rm ini}_h= 
g^{\rm ini}_h-\bar{g}^{\rm ini}\,,\hspace*{2cm}
\Delta b^{\rm ini}_h=b^{\rm ini}_h-\bar{b}^{\rm ini}\,.
\end{equation}
In other words, $\Delta g^{\rm ini}_h$ and $\Delta b^{\rm ini}_h$ 
define the
features of the initial amplitude functions $g^{\rm ini}_h$
and $b^{\rm ini}_h$ relative to their averages $\bar{g}^{\rm ini}$ and
$\bar{b}^{\rm ini}$.  Biham {\it et.\ al.}~\cite{Biham1999} have shown that the change of the
amplitudes is essentially determined by the change of the 
averages:
\begin{eqnarray} \label{BihamEquations}
g^{\rm fin}_h&=&\bar{g}^{\rm fin}+\Delta g^{\rm ini}_h \cr
b^{\rm fin}_h&=&\bar{b}^{\rm fin}+(-1)^t\Delta b^{\rm ini}_h\,,
\end{eqnarray}
where the averages $\bar{g}^{\rm fin}$ 
and $\bar{b}^{\rm fin}$ are given as follows. 
Define
\begin{eqnarray}
\omega&=&\arccos\left(1-\frac{2M}{N}\right)\;,\\
\alpha
&=&\sqrt{|\bar{b}^{\rm ini}|^2+|\bar{g}^{\rm ini}|^2\frac{M}{N-M}}\;,
   \label{eq:alpha}\\
\phi
&=&\arctan\left(\frac{\bar{g}^{\rm ini}}{\bar{b}^{\rm ini}}
    \sqrt{\frac{M}{N-M}}\,\right) \label{phi} \;.
\end{eqnarray} 
The averages are given by
\begin{eqnarray} \label{FinalAverages}
\bar{g}^{\rm fin}&=&\sqrt{\frac{N-M}{M}}\;\alpha
\,\sin(\omega t+\phi)\,, \cr
\bar{b}^{\rm fin}&=&\alpha\,\cos(\omega t+\phi)\,.
\end{eqnarray}
Let us also define the separation of the
averages
\begin{equation}   \label{separation}
\Delta^{\rm ini}=\bar{g}^{\rm ini}-\bar{b}^{\rm ini}\,,
\ \ \ \ \ 
\Delta^{\rm fin}=\bar{g}^{\rm fin}-\bar{b}^{\rm fin}\,.
\end{equation}
Equations~(\ref{BihamEquations}) imply that
after $t$ applications of the Grover operator
the values of individual amplitudes $g^{\rm fin}_h$ 
of accepted hypotheses do not change relative 
to their average~$\bar{g}^{\rm fin}$
\begin{equation} \label{FeaturePreservation}
g^{\rm fin}_h-\bar{g}^{\rm fin}=\Delta g^{\rm ini}_h
= g^{\rm ini}_h-\bar{g}^{\rm ini}\,.
\end{equation}  
The same is true for the rejected hypotheses
if $\Delta b^{\rm ini}=0$ or $t$ is even.
This observation suggests that the action
of any algorithm of the type
\begin{equation} \label{EvenAlgorithm}
{\cal A}_{aabbcc\dots}=\dots\hat{G}^2(O_c)\hat{G}^2(O_b)\hat{G}^2(O_a)
\end{equation}
can be analyzed by looking at the changes of
the average amplitudes of the accepted and
rejected hypotheses relative to various oracles
$\hat{O}_a,\hat{O}_b,\dots$. Before we proceed with
this analysis let us first clarify what kind of oracles
are relevant to this problem. 

Let $f$ be a real-valued step function which takes only three values,
\begin{equation}
f(h)=\left\{\begin{array}{ll}
             f_1 & {\rm if\ } O_{k}(h)=1\,, \cr
                         f_2 & {\rm if\ } O_{k-1}(h)=1\,,
                                          {\rm \ but\ } O_{k}(h)=0\,,\cr
                         f_3 & {\rm if\ } O_{k-1}(h)=0\,.
            \end{array}
         \right.
\end{equation}
Let us also require that $\sum_{h=0}^{N-1}f^2(h)=1$,
and introduce a quantum state
$|\Psi(f)\rangle$ that is defined by $f$ 
in a natural way,
\begin{equation} \label{f_state}
|\Psi(f)\rangle=\sum_{h=0}^{N-1}f(h)|h\rangle\,.
\end{equation}
Evidently, both $|\Psi_k\rangle$ and $|\Psi_{k-1}\rangle$
can be written in this way. 
It follows from the above discussion that the action of 
the operator $\hat{G}^2(O_k)$ on the 
state~(\ref{f_state}) can be completely described
by the changes of $f_1$, $f_2$ and $f_3$. 
Moreover $\hat{G}^2(O_k)$ preserves the value
of $\delta_{k-1}=f_2-f_3$. Similarly, the action
of $\hat{G}^2(O_{k-1})$ on (\ref{f_state})
can be completely described by the changes
of $f_1$, $f_2$ and $f_3$, and it preserves
the value of $\delta_k=f_1-f_2$.
In general, for any oracle $O$, the 
corresponding operator $\hat{G}^2(O)$
preserves the amplitude differences 
between any two hypotheses
for as long as either both hypotheses are
accepted or both are rejected by $O$.
However, if $f_{\rm good}$ and
$f_{\rm bad}$ denote the amplitudes
of an accepted and a rejected hypothesis,
respectively, then the difference
$\delta=f_{\rm good}-f_{\rm bad}$
is changed by an amount $\Delta$
which satisfies the 
inequality
\begin{equation} \label{OneStepBound}
|\Delta|
\leq\frac{4\sqrt{2}}{\sqrt{N-M}}\;,
\end{equation}
where $M$ is the number of accepted
hypotheses with respect to $O$ (see the Appendix).
Using this inequality the action of any algorithm
of the type~(\ref{EvenAlgorithm}) 
can be analyzed by calculating how individual
changes of $\delta_k$ and $\delta_{k-1}$ accumulate
during the action of the algorithm.
In order to convert 
$|\Psi_{k-1}\rangle$ into $|\Psi_{k}\rangle$
the net result of such changes must be sufficient to increase $\delta_k$ from $0$ 
to $1/\sqrt{M_k}$ and decrease
$\delta_{k-1}$ from $1/\sqrt{M_{k-1}}$ to~$0$.

It follows that all oracles that are relevant for
this task can be obtained from
$O_k$ and $O_{k-1}$.
Since each oracle is completely characterized by
the set of acceptable hypotheses, the relevant
family of oracles generated by $O_k$ and $O_{k-1}$
can be written out as oracles that correspond to
the sets
\begin{equation} \label{family}
\Omega_k,\ \Omega_{k-1},\ \Omega_{k-1}\cap\bar{\Omega}_k\,,\
{\rm and\ the\ complementary\ }
\bar{\Omega}_k,\ \bar{\Omega}_{k-1}\,,\ 
\Omega_k\cup\bar{\Omega}_{k-1}\,.
\end{equation}
Let us consider the first three oracles
from the family defined by the sets~(\ref{family}),
namely the oracles that accept hypotheses
from the sets 
\begin{equation} \label{NormalSets}
\Omega_k,\ \Omega_{k-1},\ \Omega_{k-1}\cap\bar{\Omega}_k\,.
\end{equation}
Oracles that correspond to the complementary 
sets in~(\ref{family}) can be analyzed in a 
completely analogous manner. 
The oracles corresponding to the sets (\ref{NormalSets})
are $O_k$, $O_{k-1}$ and $O_{k-1}^{k}$,
where
\begin{equation}
O_{k-1}^{k}(h)=\left\{\begin{array}{ll}
                         1& {\rm if\ } h\in\Omega_{k-1}\cap\bar{\Omega}_k\cr
                                 0& {\rm otherwise}\,.
                      \end{array}
               \right.
\end{equation}

Using the inequality~(\ref{OneStepBound}) we see that, 
regardless of its position in the algorithm, 
the operator $\hat{G}^2(O_k)$ changes 
$\delta_k$ by at most 
$d_1=4\sqrt{2}/\sqrt{N-M_k}$ without changing 
$\delta_{k-1}$.
Similarly, the operator $\hat{G}^2(O_{k-1})$ changes
$\delta_{k-1}$ by at most $d_2=4\sqrt{2}/\sqrt{N-M_{k-1}}$
without affecting the value of $\delta_k$. 
The Grover operator with the combined oracle
$\hat{G}(O_{k-1}^k)$ changes both $\delta_k$
and $\delta_{k-1}$ by an equal amount that
does not exceed $d_3=4\sqrt{2}/\sqrt{N-(M_{k-1}-M_k)}$.

Let $N(O)$ be the number of times that the oracle
$O$ is called within the algorithm. We would like
to find a lower bound on the total number of oracle
calls, $N_{\rm total}=N(O_k)+N(O_{k-1})+N(O_{k-1}^k)$,
that is needed by the algorithm to convert
$|\Psi_{k-1}\rangle$ into $|\Psi_k\rangle$.
Let $x$, $y$ and $z$ be 
the number of times that the operators $\hat{G}^2(O_{k})$,
$\hat{G}^2(O_{k-1})$ and $\hat{G}^2(O_{k-1}^k)$ appear 
in the algorithm. Then $N_{\rm total}$ is bounded from
below by the minimal value of $2(x+y+z)$ subject to the
constraints
\begin{eqnarray}
x d_1+z d_3&\geq& U\,,\cr
y d_2+z d_3&\geq& V\,,
\end{eqnarray}
where $U=1/\sqrt{M_k}$ and $V=1/\sqrt{M_{k-1}}$ are the
required changes of $\delta_k$ and $\delta_{k-1}$ respectively. 
This is a simple linear optimization problem that
should be considered in the nonnegative $x,y,z$ octant.
Keeping $M_k$ and $M_{k-1}$ constant
we obtain that in the limit of large $N$
the value of $N_{\rm total}$ is approaching
$({\sqrt{2}}/{4})\sqrt{{N}/{M_k}}\;$.

For any algorithm ${\cal A}_{abc\dots}$ the action of the
corresponding sequence of Grover operators can be rewritten
in the form similar to that in Eq.~(\ref{EvenAlgorithm}).
Using the definition of the Grover operator,
we have
\begin{equation} \label{BeginSwapping}
\hat{G}(O_{a_n})\dots\hat{G}(O_{a_3})\hat{G}(O_{a_2})\hat{G}(O_{a_1})
=
(\hat{A}\hat{O}_{a_n})
\dots
(\hat{A}\hat{O}_{a_3})(\hat{A}\hat{O}_{a_2})(\hat{A}\hat{O}_{a_1})\,,
\end{equation}
where $\hat{A}=2|\Psi_0\rangle\langle\Psi_0|-\one$.
Using the fact that $(\hat{O}_k)^2=\one$ and
$\hat{A}^2=\one$ we obtain
\begin{eqnarray}
(\hat{A}\hat{O}_{a_n})
\dots(\hat{A}\hat{O}_{a_3})
     (\hat{A}\hat{O}_{a_2})
     (\hat{A}\hat{O}_{a_1})
         &=&
\hat{A}\hat{O}_{a_n}
\dots \hat{A}\hat{O}_{a_3}
      \hat{A}\hat{O}_{a_2}
         (\hat{A}\hat{O}_{a_1})^{-1}
         (\hat{A}\hat{O}_{a_1})^2\cr
     &=&
\hat{A}\hat{O}_{a_n}
\dots \hat{A}\hat{O}_{a_3}
      \hat{A}(
          \hat{O}_{a_1a_2}\hat{A})
          (\hat{A}\hat{O}_{a_1})^2\,,
\end{eqnarray}
where $\hat{O}_{a_{1}a_{2}}=\hat{O}_{a_1}\hat{O}_{a_2}$.
Denoting $\hat{O}_{a_1a_2\dots a_j}=
          \prod_{p=1}^{j}\hat{O}_{a_p}$
we proceed
\begin{eqnarray}           \label{EndSwapping1}
\hat{A}\hat{O}_{a_n}
\dots \hat{A}\hat{O}_{a_3}
      \hat{A}(
          \hat{O}_{a_1a_2}\hat{A})
          (\hat{A}\hat{O}_{a_1})^2
     &=&
\dots \hat{A}\hat{O}_{a_3}
      \hat{A}
          (\hat{O}_{a_1a_2}\hat{A})^{-1}
          (\hat{O}_{a_1a_2}\hat{A})^2
          (\hat{A}\hat{O}_{a_1})^2\cr
     &=&
\dots \hat{A}\hat{O}_{a_4}(\hat{A}
          \hat{O}_{a_1a_2a_3})
          (\hat{O}_{a_1a_2}\hat{A})^2
          (\hat{A}\hat{O}_{a_1})^2\cr
         &=&
\hat{R}
\hat{G}^{\pm 2}(O_{a_1\dots a_n})       
\dots
          \hat{G}^2(O_{a_1a_2a_3})
          \hat{G}^{-2}(O_{a_1a_2})
          \hat{G}^2(O_{a_1})\,,\cr
          &&
\end{eqnarray}
where the $+$ and $-$ signs are chosen for odd and
even values of $n$, respectively, $O_{a_1\dots a_j}$
denote classical oracles that correspond to the 
quantum oracle $\hat{O}_{a_1\dots a_j}$,
and $\hat{R}$ is a residual operator
\begin{equation} \label{EndSwapping}
\hat{R}=\left\{\begin{array}{ll}
               \hat{G}(O_{a_1\dots a_n}) &{\rm if\ }n{\rm\ is\ even,}\cr
                           \hat{G}^{-1}(O_{a_1\dots a_n}) & {\rm otherwise}.  
               \end{array}
            \right.
\end{equation} 
Since all oracles, $\hat{O}_{a_1\dots a_j}$, belong
to the family associated with the sets~(\ref{family}),
the above arguments allow us to derive a bound on the
minimum number of oracle calls from this
family that are required to convert
$|\Psi_{k-1}\rangle$ to $|\Psi_k\rangle$.
Indeed, transformations between
Eqs.~(\ref{BeginSwapping}) and (\ref{EndSwapping})
at most double the number of oracle calls
that are used by the original algorithm.
To be more precise, if $n$ is the
number of oracle calls used by the original algorithm
(see the left-hand side of Eq.~(\ref{BeginSwapping})), then
the equivalent modified algorithm, defined by the right-hand side of 
Eq.~(\ref{EndSwapping1}), requires at most $2n+1$
oracle calls. It remains to note
that after the application of the residual
operator that concludes the modified algorithm
one has to arrive at the target state 
$|\Psi_k\rangle$, or, which is equivalent,
the algorithm
\begin{equation} \label{ActualEvenAlgorithm}
\hat{G}^{\pm 2}(O_{a_1\dots a_n})       
\dots
          \hat{G}^2(O_{a_1a_2a_3})
          \hat{G}^{-2}(O_{a_1a_2})
          \hat{G}^2(O_{a_1})
\end{equation}
must prepare the state 
$|\psi_k\rangle=G^{\pm 1}(O_{a_1\dots a_n})|\Psi_k\rangle$.
In the limit of large $N$ the state
$|\psi_k\rangle$ coincides with $|\Psi_k\rangle$.
Using an analysis analogous to that of
algorithm~(\ref{EvenAlgorithm})  we therefore
conclude that, in the limit of large $N$, the 
algorithm~(\ref{ActualEvenAlgorithm}) requires
at least $({\sqrt{2}}/{4})\sqrt{{N}/{M_k}}$
oracle calls to convert $\Psi_{k-1}$ into 
$|\psi_k\rangle$. The original algorithm, therefore,
will need, asymptotically, at least 
$({\sqrt{2}}/{8})\sqrt{{N}/{M_k}}$
oracle calls to convert $|\Psi_{k-1}\rangle$ into $|\Psi_{k}\rangle$.

\section*{Appendix}
Using the notation of 
Eqs.~(\ref{firstEquation}--\ref{FeaturePreservation}),
the inequality~(\ref{OneStepBound})
can be written as
\begin{equation} \label{AppendixEq1}
|\Delta^{\rm fin}-\Delta^{\rm ini}|
\leq\frac{4\sqrt{2}}{\sqrt{N-M}}\,,
\end{equation}
where the number of iterations is $t=2$. By definition, we have
\begin{eqnarray}
|\Delta^{\rm fin}-\Delta^{\rm ini}|
&=&\alpha\sqrt{\frac{N}{M}}\;
\Big|\,
\sin(2\omega+\xi)-\sin\xi
\,\Big|\cr
&=&\alpha \sqrt{\frac{N}{M}}\;
\left|\,2\,\sin\omega\;\cos(\omega+\xi)
\right|\cr
&\leq&4\alpha\sqrt{\frac{N}{M}}\;\left|\sin\frac{\omega}{2}\right|\;.
\end{eqnarray}
Since $0\leq\omega\leq\pi$
\begin{equation}
\sin\frac{\omega}{2}=\sqrt{\frac{1-\cos\omega}{2}}=\sqrt{\frac{M}{N}}\;,
\end{equation}
and therefore
\begin{equation} \label{DeltaAlphaBound}
|\Delta^{\rm fin}-\Delta^{\rm ini}|\leq 4\alpha\,.
\end{equation}
A bound on $\alpha$ can be obtained
using the fact that, for any $a_1,a_2,\dots,a_M\in\set{R}$
such that $\sum_{k=1}^{M}(a_k)^2=1$, we have
\begin{equation} \label{lemma}
\max\left(\frac{1}{M}\sum_{k=1}^M a_k\right)
=\frac{1}{\sqrt{M}}\;.
\end{equation} 
This can be easily shown using the method 
of Lagrange multipliers. Using~(\ref{lemma})
we obtain  
\begin{equation}
\max \bar{g}=\frac{1}{\sqrt{M}}\,,\ \ \ \ \ {\rm and }
\ \ \ \ \ \max\bar{b}=\frac{1}{\sqrt{N-M}}\,.
\end{equation}
It then follows that
\begin{equation}
\alpha < \sqrt{|\max\bar{b}|^2+|\max\bar{g}|^2\frac{M}{N-M}}
=\frac{\sqrt{2}}{\sqrt{N-M}}\;.
\end{equation}
Combining this bound with Eq.~(\ref{DeltaAlphaBound})
we obtain Eq.~(\ref{AppendixEq1}) as intended.


\begin{thebibliography}{10}

\bibitem{Bernardo1994}
J.~M.~Bernardo and A.~F.~M. Smith, {\em Bayesian Theory\/}
(Wiley, Chichester, England, 1994).

\bibitem{NielsenChuang}
M. A. Nielsen and I. L. Chuang, {\em Quantum Computation and Quantum
  Information\/} (Cambridge University Press, Cambridge, 2000).

\bibitem{Grover1997}
L.~K. Grover, 
``Quantum mechanics helps in searching for a needle in a haystack,''
Phys.\ Rev.\ Lett.\ {\bf 79},  325  (1997).

\bibitem{Biham1999}
E. Biham, O. Biham, D. Biron, M. Grassl, and D. A. Lidar, 
``Grover's Quantum Search Algorithm for an Arbitrary Initial Amplitude
    Distribution,''
Phys.\ Rev.\ A {\bf 60},  2742  (1999).

\bibitem{soksch1}
A. N. Soklakov and R. Schack, ``Efficient state preparation for a register of
quantum bits,'' {\tt quant-ph/0408045}.

\end{thebibliography}
\end{document}